# Meal-time Detection by Means of Long Periods Blood Glucose Level Monitoring via IoT Technology

Hassan M. Ahmed [1,\*], Souhail Maraoui [1], Muhammed Abd Elnaby Sadek [1], Bessam Abdulrazak [1-2], Camille Vandenberghe, [2-3], Stephen C. Cunnane[2-3], F. Guillaume Blanchet [2, 4,5,6]

[1]Ambient Intelligence Laboratory (AMI-Lab), Département d'informatique, Faculté des sciences, Université de Sherbrooke, Sherbrooke, QC J1K 2R1, Canada
[2]Research Centre on Aging, Sherbrooke, QC, J1H 4C4
[3]Brain Research Team, Université de Sherbrooke, Sherbrooke, QC, J1H4C4
[4]Département de Biologie, Faculté des sciences, Université de Sherbrooke, Sherbrooke, QC J1K 2R1, Canada
[5]Département de Mathématiques, Faculté des sciences, Université de Sherbrooke, Sherbrooke, QC J1K 2R1, Canada
[6]Département des sciences de la santé communautaire, Faculté de médecine et des sciences de la santé, Université de Sherbrooke, Sherbrooke, QC J1K 2R1, Canada

**Abstract**

Blood glucose level monitoring is of great importance, especially for participants experiencing type 1 diabetes. Accurate monitoring of their blood glucose level prevents dangerous and life-threatening situations that might be experienced by those participants. In addition, precise monitoring of blood glucose levels over long periods of time helps establishing knowledge about the daily mealtime routine which aids the medical staff to monitor participants and properly intervene in hazardous cases such as hypo- or hyper-glycemia. Establishing such knowledge will play a potential role when designing proper treatment intervention plan. In this research, we present a complete IoT framework, starting from hardware acquisition system to data analysis approaches that gives a hand for medical staff when long periods of blood glucose monitoring are essential for participants. Also, this framework is validated with real-time data collection from 7 participants over 10 successive days with temporal resolution of 5 minutes allowing for near real-time monitoring and analysis. Our results show the precisely estimated daily mealtime routines for 4 participants out of the 7 with discard of 3 participants due to huge data loss mainly. The daily mealtime routines for the 4 participants are found to be matching to have a pattern of 4 periods of blood glucose level changes corresponding to the breakfast around 8 AM, the lunch around 5 PM, the dinner around 8 PM, and finally a within-day snack around 12 PM. The research shows the potential of IoT ecosystem in support for medically related studies.

## 1. Introduction

Detecting mealtimes has a great potential to help people with diabetes to adjust glucose level as well as ensuring good metabolic health. Additionally, mealtimes and fasting intervals are important components of metabolic health [1]. Metabolic health refers to the efficient functioning of metabolic processes in the body, including how the body converts food into energy, how it uses and stores energy, and how it regulates blood sugar levels.

Following up meal habits can help managing chronical situation. For instance, factors such as dietary intake (what and how much is consumed), eating habits (food preferences and motivations, feeding practices), and environment (when, where, with whom, etc.) have a substantial impact on the emergence of chronic diseases like type 2 diabetes, heart disease, and obesity [2], [3]. Type 1 diabetes (T1D) is one of the most chronic diseases that is linked to meal habits. The International Diabetes Federation (IDF) reports that approximately 537 million individuals aged between 20 and 79 years are currently living with diabetes, and it is estimated that this number will increase to 700 million by the year 2045 [4]. Type 1 diabetes mellitus T1DM is a chronic disease that develops when the body either produces insufficient insulin or the pancreas cannot use the insulin it generates efficiently [5]. T1D patient suffers from frequent urination that leads to frequent bathroom short visits, where these short visits can be efficiently monitored using unobtrusive approaches [6] as mentioned in [7]. Insufficient insulin to decompose the generated glucose after a meal may lead to a critical health situation. Therefore, maintaining blood glucose (mainly after a meal) in normal levels is critical as blood sugar levels that are too high can lead to a variety of issues, including heart disease, kidney, nerve, and eye damage, as well as issues with wound healing [8]. In order to keep glucose level stable, Artificial Pancreas (AP) use meal data, especially for patients with T1D, to adjusts the value of insulin released based on the value measured by continuous glucose monitor (CGM)[9].

In this study we are presenting an approach for estimating the daily mealtime routine for participants based on monitoring of their blood glucose level. Using outliers' detection approach for a standard deviation sliding window gives rise to recognize the locations at which the blood glucose levels have a significant positive change, where the positive change in glucose level is considered to represent the beginning of the mealtime. Also, the probability of having a mealtime at each time period (hour) within the entire 24 hours is calculated in order to give a precise estimate for the mealtime.

The paper is organized as following, section 2 presents the related work that is concluded previously in this direction of research. Section 3 demonstrates the adopted methodology in our study, from both levels, the system level and the data analysis level. Results are presented and discussed in section 4. The study conclusion is presented in section 5.

## 2. Related work

The detection and estimate of unannounced meals have been addressed using several strategies suggested in the literature. The following categories best describe these techniques: (1) threshold-based detection using the rate of change (ROC) of glucose levels (2) outlier detection using model predictions (3) Machine learning approaches.

Palacios *et al* [1] proposed a continuous glucose monitoring (CGM), physiological monitors, and machine learning to develop an automated tool for meal detection in healthy participants. The tool utilized tree-based models such as random forests and gradient boosted trees to analyse the relationship between glucose, heart rate, physical activity, core temperature, skin temperature, and respiration rate in detecting meals. The model needed to be enhanced to accurately detect meals and select which sensor data more relevant to detecting meals.

Daniels *et al* [9] proposed a deep learning framework for meal detection and carbohydrate estimation. Long short-term memory recurrent neural network (LSTM) based on a multitask sequence-to sequence model has been used for the 20 min glucose trajectory at multiple quantiles. The proposed approach was verified on 10 adults and the result showed it could detect meals within one hour with high precision and recall except for snacks it had low recall about 24% and can detected more than one hour. The proposed model needs further improvement, by including other factors such as physiological factors.

Xu & Song [10] proposed Un-traceless Kalman filtering (UKF) to detect meals. 30 sets of volunteers were separated into three groups (adults, adolescents, and youngsters) for the simulation tests using the UVA/Padova simulator. The proposed approach has been tested on meals of six patients over two days. The results show that the UKF correctly detected 35 meals out of the 36 within one hour from start.

Mahmoudi *et al* [11] provided method for detecting meals and estimating the amount of carbohydrates in people with Type 1 Diabetes (T1D). The algorithm uses continuous glucose monitoring (CGM) data and insulin infusion rate as inputs to estimate the carbohydrate rate and detect a meal. A Kalman filter is used to augment the state-space model, and a cumulative sum algorithm and comparison with a threshold are used to announce a meal. On nine virtual T1D patients, the meal detection method and bolus calculator were tested, and the results showed 93% detection sensitivity, a median detection latency of 40 minutes, and a median bias of 5 minutes for meal beginning prediction.

Askari *et al* [12] proposed two recurrent neural network to detect meals and physical activity related to diabetic people. The proposed approaches applied on original and imputed data sets using Continuous glucose monitoring (CGM), insulin pump, and manual entries. LSTM with 1D Convolution detected "No meal or exercise" state with around 94.58% accuracy, 93.42% for identifying exercise state, and 98.05% for meals. The proposed method has a lower accuracy of 55.56% for concurrent meal-exercise events.

To guarantee that the methods are acceptable and effective in various circumstances, more research is required into the application of mealtime detection technologies in a variety of demographics, including senior individuals, those with health concerns and done on real population.

## 3. Methodology

Our methodology is based on two pillars, the underlying infrastructure pillar responsible for collecting the data, transmitting it, and storing it in our servers, and the second one is the data analysis pillar for analyzing the collected data from the participants in order to have insights about their daily mealtime routine. A complete graphical abstract for the study is presented by Figure 1.

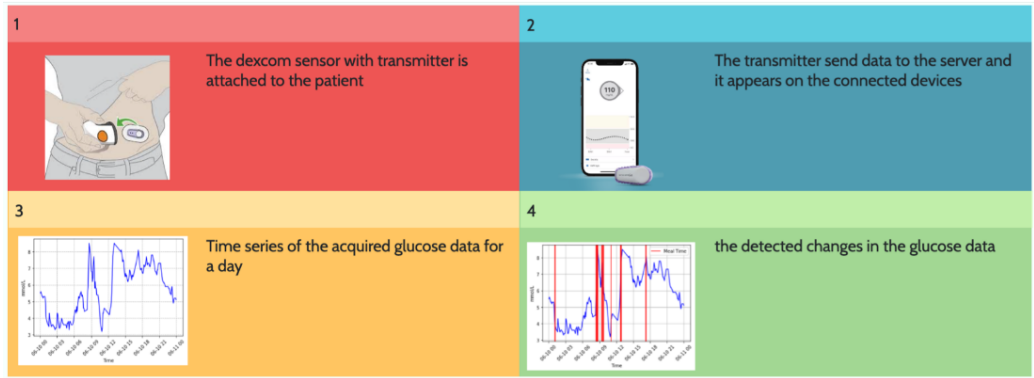

*Figure 1. General Idea Graphical Abstract.*

### 3.1. Data collection

The present study leverages AMI Platform[6], a solution developed by our team for fast and efficient deployments of IoT systems. It integrates physical elements, including sensing devices and computing hardware, as well as logical components, such as software applications and virtual computing solutions, to perform data collection, processing, transmission, storage, and analysis. In the context of this study, we have used Dexcom devices as the sensing devices for the measurement of glucose levels. They consist of a sensor that is placed on the either the participant's upper arm or abdomen to collect glucose data, and a transmitter that sends the data to a paired phone through Bluetooth. The collected glucose data is then transmitted to the cloud via a secure channel, ensuring the privacy and confidentiality of the data [13].

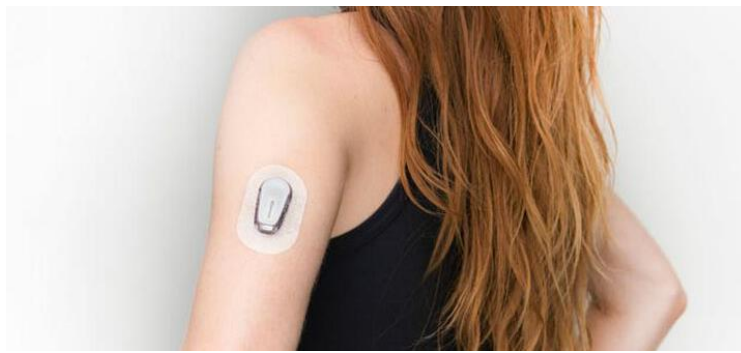

*Figure 2. Dexcom Sensor and Transmitter applied on an arm* [14].

For this study, we had 8 participants, but one dropped out before the deployment. In the end, we ran this solution for 7 participants, on June 9th, 2022, and it ended on June 19th, 2022. Each of the participants required a Dexcom sensor and transmitter, as well as a phone that hosts the Dexcom app which must be paired with so said sensor and transmitter.

This preparation process was done beforehand to ensure that all necessary equipment were available and ready for use, and to have a smoother and more convenient experience for both the participants and the nurses involved in the study. The preparation process has also enabled us to ensure that all necessary equipment was in good condition and that any issues was resolved beforehand, reducing the potential for disruptions and ensuring that the study could proceed smoothly.

One of the key benefits of our solution is its ability to collect data continuously and automatically and thus is less intrusive to the participants than traditional methods for collecting glucose data, which required frequent visits from nurses. This meant that participants were able to continue with their normal activities without interruption and reduced the burden on the nurses.

**3.2. Data analysis**

For the data analysis part, we used outliers' detection approach in order to estimate the mealtime, where the beginning of mealtime is considered to take place when the blood glucose level increases, i.e., with the rising edge of blood glucose level along the blood glucose level timeseries. To this end, we used the sliding window technique for calculating the standard deviation for each sliding window along the entire blood glucose timeseries. The temporal resolution of the study was about 5 minutes, i.e., we collected a blood glucose observation every 5 minutes for each participant, where we used a sliding window of 3 units summing up to 15 minutes window as we are interested in analyzing the short time blood glucose change.

Then we calculated the outliers for the constructed standard deviation vector using the interquartile range (IQR) approach, where based on IQR the outliers are defined to be those values falling outside 1.5 IQR below or above Q1 and Q3 respectively. A graphical representation of outliers' detection based on IQR concept is presented by Figure 3. Standard deviation shows how values are deviating around the mean value of the blood glucose level distribution within the sliding window, where having a meal will increase the blood glucose level rapidly, meaning that the rate of change of the standard deviation of the window will vary significantly if the window falls in a mealtime. Having the standard deviation vector outliers detected, we then calculated how many outliers are there per hour amongst the 24 hours of the day, where the highest number of outliers at a specific hour and/or period within the day represents the highest probability of having a meal during that time. Calculating the number of outliers for the entire 24 hours will give a close estimate about the mealtime throughout the day. The block diagram of the proposed approach for estimating the mealtime by means of blood glucose level monitoring is presented by Figure 4.

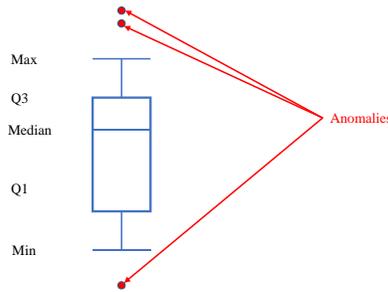

*Figure 3. Outliers' detection based on interquartile range (IQR).*

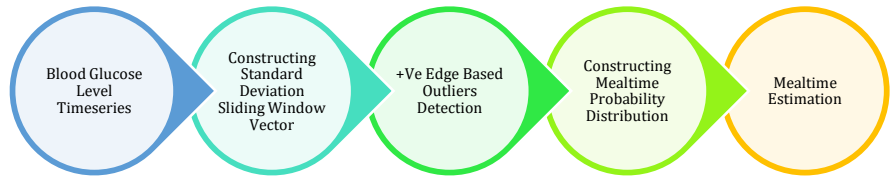

*Figure 4. Block diagram of the proposed mealtime estimation approach.*

**4. Results and discussion**

In this section we are presenting and discussing the results in a two-fold manner, the first one is from the data collection point of view, and the second one is from the data analysis point of view.

### 4.1. Data Collected

We have collected glucose data from eight participants using the Dexcom G6 sensor, which is designed to measure glucose levels every 5 minutes. We calculated the total number of glucose readings that could have been obtained from each participant based on the assumption that the sensor was functioning properly and providing a reading every 5 minutes over the 10-day period, which means that 2880 samples could have been collected over the 10 days or 240 hours of the study, starting 9$^{th}$ June noon and ending 19$^{th}$ June at noon.

The results of this analysis, which are presented in Table 1, include the number of glucose samples collected for each participant and the corresponding percentage of how many were collected out of the 2880 samples. Additionally, in Table 2 and Figure 5**Error! Reference source not found.** we see the daily samples collected per participant not including participant 2008 who dropped out.

*Table 1. Total samples collected.*

| Participant ID | Samples collected | Percentage of available data |
|---|---|---|
| H = 2008 | 0 | 0.0% |
| F = 2030 | 132 | 4.6% |
| C = 2002 | 146 | 5.1% |
| E = 2018 | 220 | 7.6% |
| A = 1014 | 2425 | 84.2% |
| B = 1026 | 2729 | 94.8% |
| G = 4008 | 2749 | 95.5% |
| D = 2011 | 2760 | 95.8% |

*Table 2. Number of daily samples per participant.*

| Participant ID | 9$^{th}$ | 10$^{th}$ | 11$^{th}$ | 12th | 13th | 14th | 15th | 16th | 17th | 18th | 19th |
|---|---|---|---|---|---|---|---|---|---|---|---|
| F = 2030 | 132 | 0 | 0 | 0 | 0 | 0 | 0 | 0 | 0 | 0 | 0 |
| C = 2002 | 81 | 65 | 0 | 0 | 0 | 0 | 0 | 0 | 0 | 0 | 0 |
| E = 2018 | 84 | 136 | 0 | 0 | 0 | 0 | 0 | 0 | 0 | 0 | 0 |
| A = 1014 | 0 | 122 | 287 | 255 | 271 | 287 | 220 | 260 | 238 | 278 | 207 |
| B = 1026 | 87 | 288 | 288 | 263 | 288 | 288 | 266 | 264 | 242 | 284 | 171 |
| G = 4008 | 91 | 284 | 288 | 266 | 282 | 288 | 270 | 264 | 263 | 286 | 167 |
| D = 2011 | 98 | 288 | 274 | 259 | 286 | 287 | 262 | 280 | 282 | 285 | 159 |

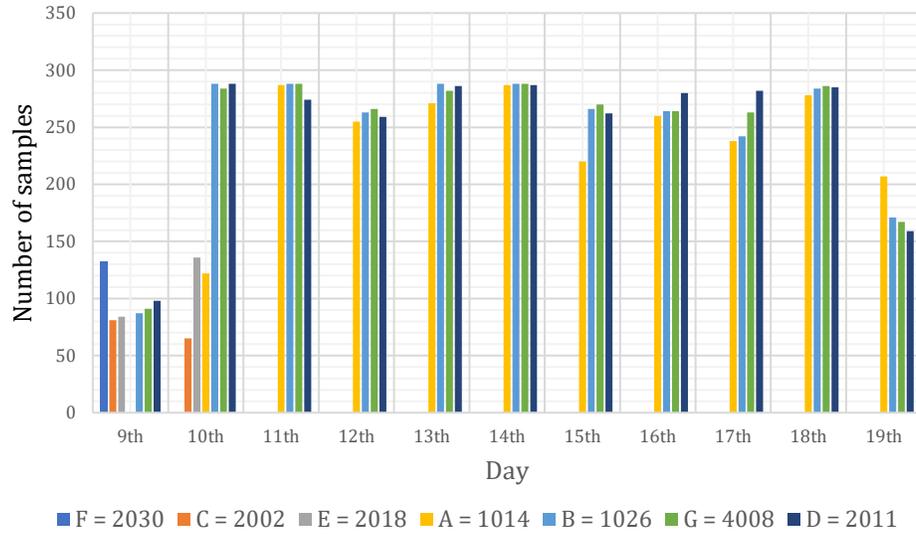

*Figure 5. Number of daily samples per participant.*

Participant 2008 dropped out of the study at the time of the deployment in fear of the sensor installation process, and participant 2030 has lost their sensor during sleep. As a result, the data collected from these participants is incomplete or entirely missing.

Given the limited data available for these participants, they will not be included in the analysis of the study results. Our analysis results regarding the estimated mealtime will focus only on the participants who had >50% glucose samples collected during the study period. Where we have enough posterior data to construct the mealtime probability over the entire 24 hours. Though in this study we did not focus on healing data acquired for other participants / participants having <50% of data, it is possible to have their blood glucose level timeseries imputed using our approach Bayesian Gaussian Process (BGaP) which can efficiently impute heavily severed timeseries with data loss at any temporal location and any missing rate >50% with the aid of posterior knowledge collected from past observations [15]. Following we are presenting the results of the data analysis part.

### 4.2. Data Analysis

Following, we first present the collected blood glucose timeseries per each participant/participant included in the study, where the timeseries is plotted as blood glucose in mmol/L versus time in day-hour-minutes format. Second, we present the boxplot with the corresponding outliers' values for each participant. It is worth mentioning here that the outliers' values at this stage represent values of standard deviation vector for each participant and not the exact monitored blood glucose values. Afterwards, we present the outliers' timestamps overlapped on the blood glucose timeseries for each participant of the 7 participants in the study. Eventually, and due to the huge amount of missing data found for participants as reported in Table 1, we considered only participants A (1014), B (1026), D (2011), and G (4008) for subsequent estimation of mealtime probability over the entire 24 hours.

**For participant A**, blood glucose values are presented over 10 days of monitoring starting from $10^{th}$ June 2022 to $19^{th}$ June 2022, with a temporal resolution of 5 minutes, where we can observe three patterns for blood glucose values, the $1^{st}$ one is at the beginning of each day with maximum blood glucose value, followed by a $2^{nd}$ one during the middle of the day with lower value, followed by a $3^{rd}$ one at the end of the day with much lower value. This pattern is observed for all the participants participating in the study, where it is clearer for participants with complete data other than those with missing data. The $1^{st}$ pattern of blood glucose values is ranging from ~12 mmol/L to ~15 mmol/L. The $2^{nd}$ pattern of values is ranging from ~10 mmol/L to ~12 mmol/L, and the $3^{rd}$ pattern of values is <10 mmol/L. This is presented by Figure 6 (a).

**For participant B**, blood glucose values are presented over 10 days of monitoring starting from 10$^{th}$ June 2022 to 19$^{th}$ June 2022, with a temporal resolution of 5 minutes. We also can observe three patterns for blood glucose values, the 1$^{st}$ one is at the beginning of each day with maximum blood glucose value, followed by a 2$^{nd}$ one during the middle of the day with lower value, followed by a 3$^{rd}$ one at the end of the day with much lower value. The 1$^{st}$ pattern of blood glucose values is ranging from ~10 mmol/L to ~12 mmol/L. The 2$^{nd}$ pattern of values is ranging from ~9 mmol/L to ~10 mmol/L, and the 3$^{rd}$ pattern of values is <9 mmol/L. This is shown by Figure 6 (b).

**For participant D**, blood glucose values are presented over 10 days of monitoring starting from 10$^{th}$ June 2022 to 19$^{th}$ June 2022, with a temporal resolution of 5 minutes. We can observe three patterns for blood glucose values, the 1$^{st}$ one is at the beginning of each day with maximum blood glucose value, followed by a 2$^{nd}$ one during the middle of the day with lower value, followed by a 3$^{rd}$ one at the end of the day with much lower value as well. The 1$^{st}$ pattern of blood glucose values is ranging from ~12 mmol/L to ~15 mmol/L. The 2$^{nd}$ pattern of values is ranging from ~10 mmol/L to ~12 mmol/L, and the 3$^{rd}$ pattern of values is <10 mmol/L. This is illustrated by Figure 7 (a).

**For participant G**, blood glucose values are presented over 10 days of monitoring starting from 10$^{th}$ June 2022 to 19$^{th}$ June 2022, with a temporal resolution of 5 minutes. Again, we can observe three patterns for blood glucose values, the 1$^{st}$ one is at the beginning of each day with maximum blood glucose value, followed by a 2$^{nd}$ one during the middle of the day with lower value, followed by a 3$^{rd}$ one at the end of the day with much lower value. The 1$^{st}$ pattern of blood glucose values is ranging from ~9 mmol/L to ~10 mmol/L. The 2$^{nd}$ pattern of values is ranging from ~8 mmol/L to ~9 mmol/L, and the 3$^{rd}$ pattern of values is <8 mmol/L. This is observed by Figure 7 (c).

We are not completely discussing the other participants due to the huge missing data experienced during their monitoring, we will be only presenting their obtained results as is.

We are starting by **participant C**, where the monitoring extended from 9$^{th}$ June 2022 in the evening till 10$^{th}$ June 2022 in the early morning. The maximum blood glucose value is ~13 mmol/L and the minimum blood glucose value is ~7 mmol/L. It is observed that blood glucose level has suddenly increased from ~7 mmol/L to ~13 mmol/L around 6 o'clock in the evening and around 8 o'clock at night respectively, indicating that the participant has got a meal at 8 o'clock. This is presented by Figure 6 (c).

**For participant E**, the monitoring extended from 9$^{th}$ June 2022 in the evening till 10$^{th}$ June 2022 in the morning. The blood glucose level is showing a decaying behavior through whole night where it is ~12 mmol/L around 6 o'clock in the evening and ~6 mmol/L around 5 o'clock in the morning. Also, it is clearly observed from the time series of that participant that there was a meal taken around 9 o'clock in the morning, where his blood glucose level has risen up rapidly from ~7 mmol/L to ~12 mmol/L at its peak indicating the probable mealtime to be around that time. This is shown by Figure 7 (b).

**For participant F**, we could not acquire data for him except for 6 hours, and we do not consider his data to be precise, hence we rejected him from the analysis.

For outliers' detection, the boxplot results for each participant along with the outliers as red dots are presented in Figure 8. We can observe that the average IQR for all participants is below 1, while the outliers' values are below 2 except for participants A and D where their corresponding outliers' values are ~5 and ~3 respectively. Figure 9 and Figure 10 are representing the detected blood glucose level outliers as vertical red lines overlapped on their corresponding timeseries. Each red line represents positive change in the blood glucose level. Also, it can be seen that for each day there are 3 or 4 groups of detected blood glucose changes representing the 3 main meals during the day, and the 4$^{th}$ group if found represents that the participant had a snack at that time.

The estimated daily mealtime routines (discrete mealtime probability) for each participant are presented in Figure 11 and Figure 12 respectively, where each bar in the mealtime routine represents the probability of having the meal at that time. From these figures we can conclude that all participants had breakfast between 8 AM to 9 AM in the morning, the lunch at 5 PM, and the dinner between 7 PM to 8 PM at night. It is observed the four participants had a small meal or a snack a 12 PM in the noon. From the estimated discrete mealtime probability plot of participants D and G we can conclude that both of them has a snack during the night between 3 AM to 4 AM.

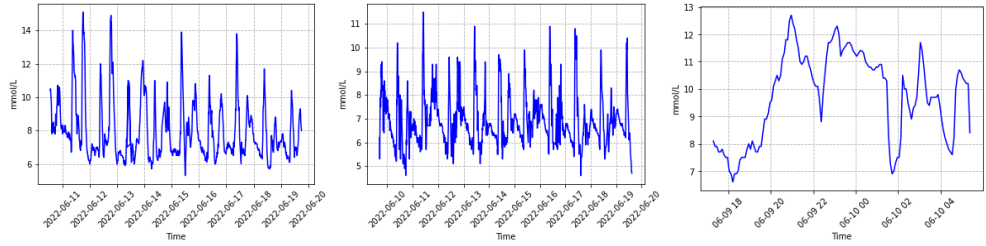

*(a)*     *(b)*     *(c)*

*Figure 6. Blood glucose level for a) participant A, b) participant B, and c) participant C.*

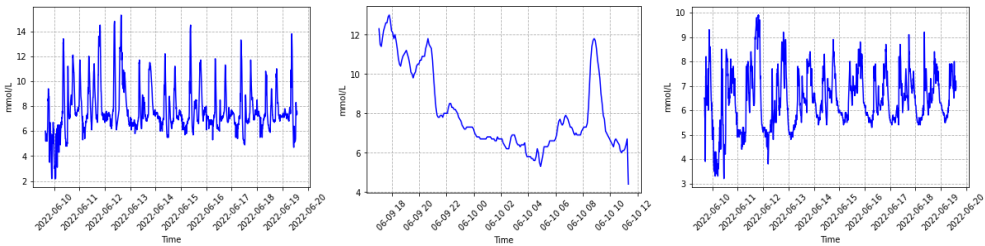

*(a)*     *(b)*     *(c)*

*Figure 7. Blood glucose level for a) participant D, b) participant E, and c) participant G.*

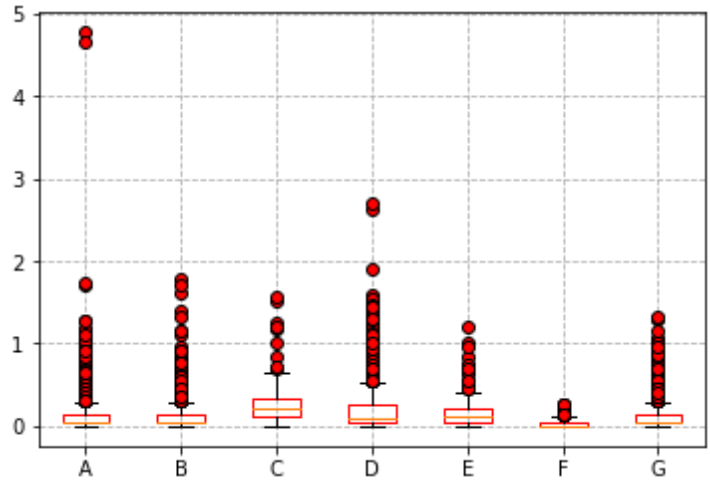

*Figure 8. Boxplot for each participant of the 7 participants from A to G.*

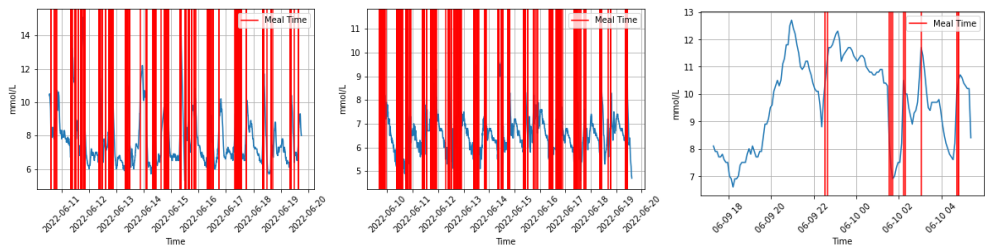

*(a)*     *(b)*     *(c)*

*Figure 9. Detected Blood Glucose Level Outliers as Mealtime for a) participant A, b) participant B, and c) participant C.*

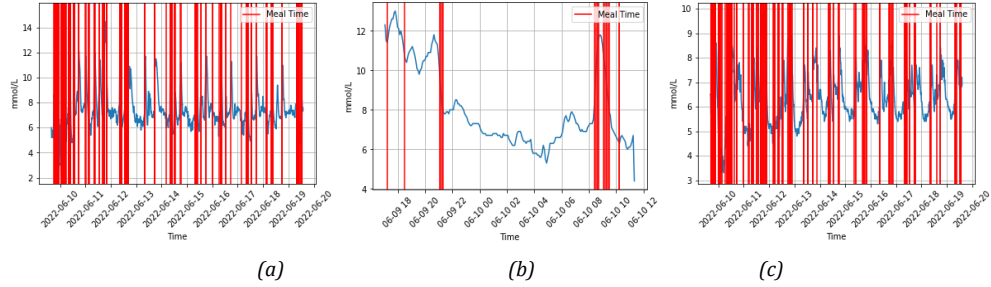

*(a)* *(b)* *(c)*
*Figure 10. Detected Blood Glucose Level Outliers as Mealtime for a) participant D, b) participant E, and c) participant G.*

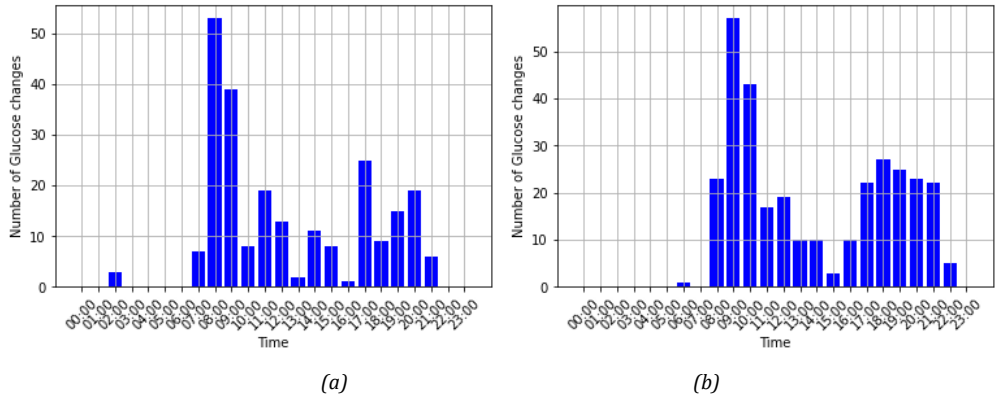

*(a)* *(b)*
*Figure 11. Estimated daily mealtime routine chart for a) participant A, and b) participant B.*

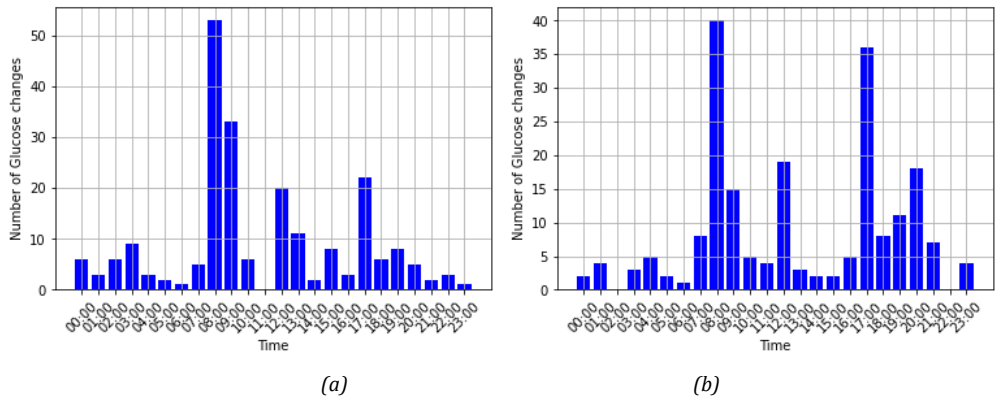

*(a)* *(b)*
*Figure 12. Estimated daily mealtime routine chart for a) participant D, and b) participant G.*

## 5. Conclusion

We have presented in this paper an approach for meal-time detection based on discrete mealtime probability estimation by means of long periods blood glucose level monitoring via IoT technology. We have monitored blood glucose level for 7 participants for 10 consecutive days. We used Dexcom devices for each participant with a temporal resolution of 5 minutes to construct blood glucose level timeseries for each participant. Our research included two parts, the system implementation part and the data analysis part. The system implementation is concerned about collecting blood glucose level data from the participants, while the data analysis part is concerned about timeseries data analysis for deducting the daily mealtime routines for each participant.

The present study has demonstrated the use of the AMI Platform for an efficient deployment of IoT systems in the context of monitoring glucose levels in participants using Dexcom devices. The solution was able to collect data continuously and automatically, which reduced the burden on the participants and nurses involved in the study. Four participants out of the 7 participants had a high percentage of data collected out of the total possible samples under ideal conditions, with participant D having the highest at 95.8%, followed by participant G with 95.5%, and participants B and A with 94.8% and 84.2% respectively. However, one participant dropped out on the day of the deployment, and another lost their sensor during sleep, highlighting the importance of participant cooperation for successful data collection in studies using IoT solutions.

Furthermore, we have estimated the daily mealtime time for 4 participants from the 7 participating participants. The others are discarded from analysis due to the huge amount of missing data found in their corresponding blood glucose level time series. For the 4 analyzed participants we were able to precisely estimate their daily mealtime routine thanks to the outliers' detection approach. Outliers' detection approach is applied on a standard deviation sliding window for each participant to detect the locations at which their blood glucose level's rate of change is positively significant, where those locations represent the beginning of the mealtime. The daily mealtime routine is found to follow a pattern of 3 or 4 periods at which there is a possible meal taken by each participant. Those periods represent breakfast around 8 AM to 9 AM, lunch around 5 PM, and dinner meals around 7 PM to 8PM, where there is a snack taken at roughly 12 PM respectively.

In general, we conclude that is possible to detect un-announced mealtime for a person, where this can help diabetes suffering patient for getting a suitable alert in case of hyper- or hypo-glycemia, during such situations their lives are considered in threat. Also, it is possible for medical staff to monitor multiple participants on a real-time basis and establishing valuable information about their daily mealtime routine, which helps the medical staff to give proper treatment relying on solid past knowledge other than the participant's memory, which may be weak in multiple situations. In addition, it is possible after this study to establish a precise range of normal blood glucose levels for a population in order to study the effect of any proposed cure for diabetes participants in the future.